\documentclass[11pt,a4paper,english,superscriptaddress,aps,preprint]{revtex4}
\usepackage{xcolor}
\usepackage{amsfonts}
\usepackage{graphics}
\usepackage{graphicx}
\usepackage{hyperref}
\usepackage{slashed}
\usepackage{amsmath}
\usepackage{amssymb}
\makeatletter
\usepackage{babel}
\newcommand{\bea}{\begin{eqnarray}}
\newcommand{\eea}{\end{eqnarray}}

\begin{document}

\title{Effective Potential in the 3D Massive 2-form Gauge Superfield Theory}

\author{F. S. Gama}
\email{fisicofabricio@yahoo.com.br}
\affiliation{Departamento de F\'{\i}sica, Universidade Federal da Para\'{\i}ba\\
 Caixa Postal 5008, 58051-970, Jo\~ao Pessoa, Para\'{\i}ba, Brazil}

\author{J. R. Nascimento}
\email{jroberto@fisica.ufpb.br}
\affiliation{Departamento de F\'{\i}sica, Universidade Federal da Para\'{\i}ba\\
 Caixa Postal 5008, 58051-970, Jo\~ao Pessoa, Para\'{\i}ba, Brazil}

\author{A. Yu. Petrov}
\email{petrov@fisica.ufpb.br}
\affiliation{Departamento de F\'{\i}sica, Universidade Federal da Para\'{\i}ba\\
 Caixa Postal 5008, 58051-970, Jo\~ao Pessoa, Para\'{\i}ba, Brazil}

\author{P. J. Porf\'{i}rio}
\email{pporfirio89@gmail.com}
\affiliation{Departament of Physics and Astronomy, University of Pennsylvania, Philadelphia, PA 19104, USA}

\begin{abstract}
In the $\mathcal{N}=1$, $d=3$ superspace, we propose a massive superfield theory formulated in terms of a spinor gauge superfield, whose component content includes a two-form field, and a real scalar matter superfield. For this model, we explicitly calculate the one-loop correction to the superfield effective potential. In particular, we show that the one-loop effective potential is independent the gauge-fixing parameters.
\end{abstract}

\maketitle

\section{Introduction}

As it is well known, the most studied supersymmetric models are based on a gauge multiplet, describing gauge fields and their superpartners, and a scalar multiplet describing the usual matter. Many issues related to these models in different cases were studied,  both in classical and quantum contexts. Nevertheless, other supersymmetry multiplets, including those ones presented in \cite{SGRS}, also deserve to be considered. One of the important examples is the tensor multiplet whose component content includes an antisymmetric tensor field \cite{Siegel}, which, as is well known, plays an important role since it emerges in string theory \cite{String}, and it has been studied in many other contexts, such as Lorentz symmetry violation \cite{Lorentz}, quantum equivalence \cite{Equivalence,Equivalence0,Equivalence1},  paramagnetism-ferromagnetism phase transition \cite{Transition}, and cosmological inflation \cite{Inflation}. The quantum impacts of the tensor multiplet were studied in the four-dimensional space-time, where it is described by the chiral spinor superfield, in \cite{ourtensor}, where the one-loop effective potential was calculated in the model including this superfield some further development of this model has been carried out in \cite{Almeida1}. Therefore, the natural problem consists in generalization of this study to three dimensions through treating a theory of the three-dimensional tensor multiplet which is known to be described by the gauge spinor superfield. The corresponding superfield description of a theory on the tree level has been developed already in \cite{Almeida}. Therefore, it is natural to promote this study to the quantum level, introducing a coupling of the gauge spinor superfield to some matter, and calculating the one-loop quantum corrections in this theory. This is the aim we pursue in this paper.

Our calculations are based on the methodology of calculating the superfield effective potential developed for the three-dimensional case originally in \cite{ourEP} and then used for various three-dimensional superfield theories in a number of papers, f.e. \cite{GNP,two-loop}. We calculate the effective potential in the one-loop approximation.

The structure of our paper looks like follows. In the section 2 we consider the classical actions of a theory involving the  real spinor superfield. In the section 3, we explicitly calculate the one-loop effective potential for this theory, and in the section 4, the results are summarized.


\section{The Model}
\label{model}
By imposing some constraints on the field strength for the three-dimensional 2-form gauge superfield $\Gamma_{AB}$, it is possible to show that $\Gamma_{AB}$ can be completely expressed in terms of a prepotential $B_\alpha$, which is an unconstrained real spinor gauge superfield \cite{SGRS}. Having this in mind, we start with the following definition
\bea
\label{kinetic}
S_k[B_\alpha,\Phi]=-\frac{1}{2}\int d^5z \left[\left(D_\alpha G\right)^2+\left(D_\alpha\Phi\right)^2\right] \ ,
\eea
where $G\equiv-D^\alpha B_\alpha$ is a gauge invariant field strength and $\Phi$ is the usual real scalar matter superfield. The identity $D^\alpha D^\beta D_\alpha=0$ ensures that $S_k$ is invariant under the gauge transformation
\bea
\label{gaugetrans1}
\Phi\rightarrow\Phi^\Lambda=\Phi \ ; \
B_\alpha\rightarrow B_\alpha^\Lambda=B_\alpha+\frac{1}{2}D^\beta D_\alpha\Lambda_\beta \ ,
\eea
with a spinor gauge parameter $\Lambda_\alpha$. The model (\ref{kinetic}) is an example of first-stage reducible theory. 
Indeed, the parameter $\Lambda_\alpha$ in (\ref{gaugetrans1}) is not unique, but it is defined up to  the transformation $\Lambda^\prime_\beta=\Lambda_\beta+D_\beta L$, where $L$ is an arbitrary scalar superfield, 
in other words, there are gauge transformations for gauge parameters. The methodology for studying reducible theories has been developed in \cite{BV}, and the general discussion of such theories can be found in \cite{GPS}. The four-dimensional analogue of the theory (\ref{kinetic}), within supergravity context, has been considered in \cite{Equivalence0}. 

Now, we want to introduce mass terms for the theory (\ref{kinetic}). These terms are defined as
\bea
\label{mass}
S_m[B_\alpha,\Phi]=\int d^5z \Big[m\Phi G+m_B^2 B^\alpha B_\alpha+\frac{1}{2}m_\Phi\Phi^2\Big] \ .
\eea
The term $m\Phi G$ corresponds to the supersymmetric extension of the topological BF model \cite{Almeida}. It is worth to note that  $m_B^2 B^\alpha B_\alpha$ explicitly breaks the gauge invariance of $S_m$ under the transformation (\ref{gaugetrans1}).

Let us check that $S_k+S_m$ indeed describes a massive gauge
theory. For this, we need to obtain the free superfield equations for $B_\alpha$ and $\Phi$, which are derived from the principle of stationary action. Thus, we get from $S_k+S_m$:
\bea
\label{1}
\frac{\delta(S_k+S_m)}{\delta B^\alpha}&=&D_\alpha D^2G+mD_\alpha\Phi+2m^2_B B_\alpha=0;\\
\label{2}
\frac{\delta(S_k+S_m)}{\delta\Phi}&=&D^2\Phi+mG+m_\Phi\Phi=0.
\eea
On the one hand, if $m_B=m_\Phi=0$ and $m\neq0$, then we can multiply Eq. (\ref{1}) by $D^\alpha/2$ and use $(D^2)^2=\Box$ to obtain 
\bea
\left(\Box-m^2\right)G=0 \ .
\eea
We can carry out a similar calculation to show that $D_\alpha\Phi$ satisfies a Klein-Gordon equation.

On the other hand, if $m_B\neq0$ and $m=0$, then we can multiply Eq. (\ref{1}) by $D^2D^\alpha D^\gamma$ and use $D^\alpha D^\gamma D_\alpha=0$ to obtain 
\bea
D^2D^\alpha D^\gamma B_\alpha&=&0.
\eea
Substituting this back into the equation (\ref{1}) and using $D^2[D_\alpha,D_\beta]=2C_{\beta\alpha}\Box$, we get
\bea
\left(\Box-m_B^2\right)B_\alpha=0.
\eea
It is trivial to show that $\Phi$ also satisfies a Klein-Gordon equation for $m_\Phi\neq0$ and $m=0$. Therefore, we demonstrated that $S_k+S_m$ describes a massive gauge theory.

Since the model under investigation $S_k+S_m$ is a free superfield theory, and the main purpose of this paper is to calculate the one-loop effective potential, then we must to extend $S_k+S_m$ to include interactions. Here, we define the interaction between $B_\alpha$ and $\Phi$ as
\bea
\label{int}
S_{int}[B_\alpha,\Phi]=\int d^5z\Big[V_0(\Phi)+V_1(\Phi)G+\frac{1}{2}V_2(\Phi)G^2+V_3(\Phi)B^\alpha B_\alpha\Big],
\eea
where $V_i(\Phi)$'s are analytical functions of their arguments. Note that we have ignored in (\ref{int}) terms higher than quadratic in $B_\alpha$ due to the fact that these terms will not contribute at the one-loop level to the effective potential. Moreover, in addition to (\ref{mass}), $S_{int}$ also lacks gauge invariance. 

The lack of gauge invariance of $S_{m}$ and $S_{int}$ is inconvenient for quantum calculations. In order to improve the situation, we will restore the gauge symmetry by introducing a St\"{u}ckelberg superfield $\Omega_\alpha$ \cite{RR}. Thus, instead of the theory $S_k+S_m+S_{int}$, we will study in this work the following gauge-invariant theory, obtained from the previous one through adding some new terms, whose action is
\bea
\label{total}
S[B_\alpha,\Omega_\alpha,\Phi]&=&\frac{1}{2}\int d^5z\bigg\{GD^2G+\Phi D^2\Phi+m_\Phi\Phi^2+2V_0(\Phi)+2\big[m\Phi+V_1(\Phi)\big]G+V_2(\Phi)G^2\nonumber\\
&+&2\big[m_B^2+V_3(\Phi)\big]\left(B^\alpha-\frac{W^\alpha}{m_B}\right)\left(B_\alpha-\frac{W_\alpha}{m_B}\right)\bigg\},
\eea
with $W_\alpha\equiv\frac{1}{2}D^\beta D_\alpha\Omega_\beta$. The new action (\ref{total}) is invariant under the following transformations
\bea
\Phi\rightarrow\Phi^\Lambda=\Phi \ ; \
B_\alpha\rightarrow B_\alpha^\Lambda=B_\alpha+\frac{1}{2}D^\beta D_\alpha\Lambda_\beta \ ; \ \Omega_\alpha\rightarrow \Omega_\alpha^\Lambda=\Omega_\alpha+m_B\Lambda_\alpha,
\eea
with spinor gauge parameter $\Lambda_\alpha$. Moreover, $S$ is also invariant under the gauge transformation
\bea
\Phi\rightarrow\Phi^K=\Phi \ ; \
B_\alpha\rightarrow B_\alpha^K=B_\alpha \ ; \ \Omega_\alpha\rightarrow \Omega_\alpha^K=\Omega_\alpha+D_\alpha K \ ,
\eea
with an arbitrary scalar gauge parameter $K$.

Since (\ref{total}) is gauge invariant, it follows that a gauge fixing is necessary for the calculation of quantum corrections to the effective potential. Thus, the gauge-fixed action is defined as the sum $S+S_{gf}$, where $S$ is given in Eq. (\ref{total}) and the gauge-fixing term $S_{gf}$ is given by
\bea
S_{gf}[B_\alpha,\Omega_\alpha]=\frac{1}{2}\int d^5z\left(\begin{array}{cc}
B^\alpha & \Omega^\alpha
\end{array}\right)
\left(\begin{array}{cc}
\displaystyle-\frac{1}{\alpha}D^2D^\beta D_\alpha & \displaystyle m_BD^\beta D_\alpha \\
\displaystyle m_BD^\beta D_\alpha & \displaystyle-2\alpha m_B^2{\delta_\alpha}^\beta-\frac{1}{\xi}D^2D_\alpha D^\beta
\end{array}\right)
\left(\begin{array}{c}
B_\beta \\
\Omega_\beta
\end{array}\right).
\eea
In particular, if we choose $m=0$ and the supersymmetric Fermi-Feynman gauge $\alpha=\xi=1$, the kinetic terms take the particularly simple forms $\sim B^\alpha(\Box-m_B^2)B_\alpha$ and $\sim\Omega^\alpha(\Box-m_B^2)\Omega_\alpha$. 

Of course, there be also ghosts in the gauge-fixed action. Indeed, besides the usual ghosts, there are also ghosts for ghosts due to the fact that (\ref{total}) describes a first-stage reducible theory. However, since the ghosts do not interact with the scalar superfield $\Phi$, it follows that the ghost terms do not contribute to the one-loop effective potential. For this reason, we can omit such terms. We note that the similar situation takes place in four dimensions \cite{Almeida1}.

\section{One-loop Calculations}
\label{OLC}

In this section, we calculate the one-loop effective potential for the  theory (\ref{total}). To do this, we employ the background field method \cite{BOS}. Within this approach, we perform the calculations by making a linear split of the superfields into background superfields $(B_\alpha,\Omega_\alpha,\Phi)$ and quantum fluctuations $(b_\alpha,\omega_\alpha,\phi)$:
\bea
\label{split}
B_\alpha\rightarrow B_\alpha+b_\alpha \ ; \ \Omega_\alpha\rightarrow\Omega_\alpha+\omega_\alpha \ ; \ \Phi\rightarrow\Phi+\phi.
\eea
By definition, the effective potential depends only on the matter superfield $\Phi$. Thus, we assume a trivial background for the gauge superfields $B_\alpha$, $\Omega_\alpha$, and the derivatives of $\Phi$:
\bea
B_\alpha=\Omega_\alpha=0; \ D_\alpha\Phi=0; \ \partial_{\alpha\dot{\alpha}}\Phi=0,
\eea
while  a background $\Phi$ differs from zero.

For the sake of simplicity, before we consider the general problem, we first study the particular case where $m_B^2=V_3(\Phi)=0$. We denote the effective potential calculated in this case by $K^{(1)}_A$. The importance of this choice is based by the fact that in this case the superfield $\Omega_\alpha$ completely decouples from theory (\ref{total}). Therefore, expanding $S+S_{gf}$ around
the background superfields and keeping only the quadratic terms in the quantum fluctuations, one finds
\bea
&&S_2[\Phi;\phi,b_\alpha]=S_K+S_{INT};\\
&&S_K=\frac{1}{2}\int d^5z\Big\{b^\alpha\big[D^2(D_\alpha D^\beta-\frac{1}{\alpha}D^\beta D_\alpha)\big]b_\beta+\phi D^2\phi\Big\};\\
\label{vertices}
&&S_{INT}=\frac{1}{2}\int d^5z\big[(m_\Phi+V^{\prime\prime}_0)\phi^2+2(m+V_1^\prime)\phi g+V_2g^2\big],
\eea
where $g\equiv-D^\alpha b_\alpha$, $V_1^{\prime}\equiv dV_1/d\Phi$, and $V_0^{\prime\prime}\equiv d^2V_0/d\Phi^2$.

The interaction vertices can be read off directly from $S_{INT}$, and the propagators are obtained by inverting the differential operators in $S_K$, being given by 
\bea
\label{prop1}
\langle b_\alpha(1)b^\beta(2)\rangle&=&-\frac{1}{4k^4}D_1^2\big(D_{1\alpha}D_1^\beta-\alpha D^\beta_1D_{1\alpha}\big)\delta^2(\theta_1-\theta_2) \ ;\\
\label{prop2}
\langle\phi(1)\phi(2)\rangle&=&\frac{D_1^2}{k^2}\delta^2(\theta_1-\theta_2).
\eea
Notice in Eq. (\ref{vertices}) that the quantum superfield $b_\alpha$ interacts with the background one $\Phi$ through its field strength $g$. Thus, instead of the propagator $\langle b_\alpha(1)b^\beta(2)\rangle$, it is sufficient to use the propagator with no spinor indices $\langle g(1)g(2)\rangle$, which is given by:
\bea
\label{prop3}
\langle g(1)g(2)\rangle=D_1^\alpha D_{2\beta}\langle b_\alpha(1)b^\beta(2)\rangle =\frac{D_1^2}{k^2}\delta^2(\theta_1-\theta_2).
\eea
It is clear that (\ref{prop3}) does not depend on the gauge parameter $\alpha$ introduced in the gauge-fixing procedure. Therefore, before we start the calculation of the one-loop effective potential $K^{(1)}_A(\Phi)$, we can already conclude that $K^{(1)}_A(\Phi)$ is gauge independent as it occurs in some other three-dimensional supergauge theories, see f.e. \cite{GNP}.

The propagators (\ref{prop2}), (\ref{prop3}), and the vertices (\ref{vertices}) can be written in a matrix form.  In order to do this, we make the definitions
\bea
\chi_i\equiv\left(\begin{array}{c}
g \\
\phi
\end{array}\right) \ ; \ \chi^j\equiv\left(\begin{array}{cc}
g \ & \ \phi
\end{array}\right) \ ; \ {M_i}^j\equiv\left(\begin{array}{cc}
\displaystyle V_2 & \displaystyle m+V_1^\prime \\
\displaystyle m+V_1^\prime & \displaystyle m_\Phi+V_0^{\prime\prime}
\end{array}\right) \ ,
\eea
so that we can show that
\bea
\label{matrixform}
\langle\chi_i(1)\chi^j(2)\rangle=\frac{D_1^2}{k^2}{\delta_i}^j\delta^5(\theta_1-\theta_2) \ ; \ S_{INT}=\frac{1}{2}\int d^5z\chi^i{M_i}^j\chi_j.
\eea
These propagators and vertices are quite similar to ones used in our previous work \cite{GNP}, where we have calculated $K^{(1)}(\Phi)$ for a generic superfield higher-derivative gauge theory. Due to this similarity, we simply quote the result here:
\bea
\label{integral}
K^{(1)}_A(\Phi)&=&\frac{1}{2}\int \frac{d^3k}{(2\pi)^3}\frac{1}{|k|}\left[\arctan\left(\frac{\lambda_+}{|k|}\right)+
\arctan\left(\frac{\lambda_-}{|k|}\right)\right],
\eea
where the $\lambda$'s are the eigenvalues of the matrix ${M_i}^j$, and $|k|=\sqrt{k^2}$.

Substituting the eigenvalues into (\ref{integral}) and calculating the integral over the momenta, we obtain
\bea
\label{poteff1}
K^{(1)}_A(\Phi)=-\frac{1}{16\pi}\left[\left(m_\Phi+V^{\prime\prime}_0\right)^2+2\left(m+V^{\prime}_1\right)^2+V_2^2\right].
\eea
Just as in the usual three-dimensional field theories, this one-loop contribution to the effective potential is UV finite, and its functional structure is given by a polynomial function of $V^{\prime\prime}_0,V^{\prime}_1$, and $V_2$. Indeed, in contrast to four-dimensional theories, logarithmic functions begin to occur only at the two-loop level due to the divergences of the Feynman integrals \cite{ourEP,two-loop}. Additionally, as we already said before, (\ref{poteff1}) is independent of the gauge-fixing parameter $\alpha$. This result was expected because the theory (\ref{total}) with $m_B^2=V_3(\Phi)=0$ is classically equivalent to a theory with two massive real scalar superfields, even though $G$ is by definition a field strength. However, it is not clear whether $K^{(1)}(\Phi)$ is independent of $\alpha$ when $m_B^2\neq V_3(\Phi)\neq0$. Thus, let us move on and calculate $K^{(1)}(\Phi)$ in the general case $m_B^2\neq V_3(\Phi)\neq0$. We denote the effective potential calculated in this case by $K^{(1)}_B$.

Again, in order to evaluate the $K^{(1)}_B(\Phi)$ one should expand (\ref{total}) around the background superfields (\ref{split}) and keep the terms quadratic in the fluctuations:
\bea
\label{undia}
S_2[\Phi;\phi,b_\alpha,\omega_\alpha]&=&\frac{1}{2}\int d^5z\Big\{b^\alpha\big[D^2(D_\alpha D^\beta-\frac{1}{\alpha}D^\beta D_\alpha)+2m_B^2{\delta_\alpha}^\beta\big]b_\beta+\omega^\alpha\big[D^2(D^\beta D_\alpha\nonumber\\
&-&\frac{1}{\xi}D_\alpha D^\beta)-2\alpha m_B^2{\delta_\alpha}^\beta\big]\omega_\beta+\phi D^2\phi+(m_\Phi+V^{\prime\prime}_0)\phi^2+2(m+V_1^\prime)\phi D_\alpha b^\alpha\nonumber\\
&-&V_2 b^\alpha D_\alpha D^\beta b_\beta+2V_3 b^\alpha b_\alpha-2\frac{V_3}{m_B}\omega^\beta D^\alpha D_\beta b_\alpha+\frac{V_3}{m_B^2}\omega^\alpha D^2D^\beta D_\alpha \omega_\beta\Big\},
\eea
where we have now taken into account the contributions of $\omega_\alpha$.

The quadratic mixing terms between the quantum superfields make the calculations troublesome. Fortunately, we can overcome this complication by a non-local change of variables in the path integral, as was done in \cite{non-local}. Thus, we can diagonalize (\ref{undia}) with the choice
\bea
\label{transform1}
\phi(z)&\longrightarrow&\phi(z)-\int d^5wG(z,w)\left[m+V_1^\prime\left(\Phi(w)\right)\right]D_{w\alpha} b^\alpha(w);\\
\label{transform2}
\omega_\alpha(z)&\longrightarrow&\omega_\alpha(z)+\int d^5w{G_\alpha}^\beta(z,w)\frac{V_3\left(\Phi(w)\right)}{m_B}D^\gamma_w D_{w\beta} b_\gamma(w),
\eea
where $G(z,w)$ and ${G_\alpha}^\beta(z,w)$ are Green's functions, which are defined as solutions of the equations 
\bea
&&\left(D^2+m_\Phi+V_0^{\prime\prime}\right)G(z,w)=\delta^5(z-w);\\
&&D^2\left[\left(1-\alpha\frac{m_B^2}{\Box}+\frac{V_3}{m_B}\right)D^\gamma D_\alpha+\left(\alpha\frac{m_B^2}{\Box}-\frac{1}{\xi}\right)D_\alpha D^\gamma\right]{G_\gamma}^\beta(z,w)={\delta_\alpha}^\beta\delta^5(z-w).
\eea
It is possible to show that these functions can be expressed in the form
\bea
G(z,w)&=&\frac{D^2-\left(m_\Phi+V_0^{\prime\prime}\right)}{\Box-\left(m_\Phi+V_0^{\prime\prime}\right)^2}\delta^5(z-w);\\
{G_\gamma}^\beta(z,w)&=&\frac{D^2}{4\Box}\left[\frac{1}{\Box-\alpha m^2_B+\frac{V_3}{m^2_B}\Box}D^\beta D_\gamma-\frac{\xi}{\Box-\alpha\xi m^2_B}D_\gamma D^\beta\right]\delta^5(z-w).
\eea
It is worth to point out that we assume that the quantum variable $b_\alpha$ does not change under the transformations (\ref{transform1}) and (\ref{transform2}). For this reason, these transformations correspond to translations on the field space, so that the corresponding Jacobian is equal to unity. 

Therefore, after the change of variables (\ref{transform1}) and (\ref{transform2}), the functional $S_2$ can be rewritten as:
\bea
\label{diagS2}
S_2&=&\frac{1}{2}\int d^5z\Big\{b^\alpha\left[\left(D^2\left(1-\frac{m_B^2+V_3}{\Box}\right)-V_2\right)D_\alpha D^\beta+D^2\left(\frac{m_B^2+V_3}{\Box}-\frac{1}{\alpha}\right)D^\beta D_\alpha\right]b_\beta\nonumber\\
&+&\omega^\alpha D^2\left[\left(1-\alpha\frac{m^2_B}{\Box}+\frac{V_3}{m^2_B}\right)D^\beta D_\alpha+\left(\alpha\frac{m^2_B}{\Box}-\frac{1}{\xi}\right)D_\alpha D^\beta\right]\omega_\beta+\phi\left(D^2+m_\Phi+V^{\prime\prime}_0\right)\phi\Big\}\nonumber\\
&-&\frac{1}{2}\int d^5zd^5w b^\alpha(z)b_\beta(w)\Bigg[\left(m+V_1^\prime\right)^2\frac{D^2+m_\Phi+V_0^{\prime\prime}}{\Box-\left(m_\Phi+V_0^{\prime\prime}\right)^2}D_\alpha D^\beta+\left(\frac{V_3}{m_B}\right)^2\nonumber\\
&\times&\frac{D^2D^\beta D_\alpha}{\Box-\alpha m^2_B+\frac{V_3}{m^2_B}\Box}\Bigg]\delta^5(z-w).
\eea
In principle, we could derive the  Feynman rules for the functional (\ref{diagS2}) and calculate the one-loop supergraphs which contribute to the effective potential. However, it is much easier to perform the calculation using the well-known formula for the one-loop Euclidean effective action \cite{BK,GRU}
\bea
\label{1loopEA}
\Gamma^{(1)}_B[\Phi]=-\frac{1}{2}\textrm{sTr}\ln\widehat{\mathcal O} \ , 
\eea
where $\textrm{sTr}$ denotes the supertrace over the discrete and continuous indices of $\widehat{\mathcal O}$.

It follows from (\ref{diagS2}) that $\widehat{\mathcal O}$ is a block diagonal matrix. Thus, Eq. (\ref{1loopEA}) can be split into three contributions:
\bea
\label{totalgamma}
\Gamma^{(1)}_B[\Phi]=\Gamma_\omega[\Phi]+\Gamma_b[\Phi]+\Gamma_\phi[\Phi],
\eea
where
\bea
\label{gammao}
\Gamma_\omega[\Phi]&=&\frac{1}{2}\textrm{Tr}\ln\left[\left(1-\alpha\frac{m^2_B}{\Box}+\frac{V_3}{m^2_B}\right)D^2D^\beta D_\alpha+\left(\alpha\frac{m^2_B}{\Box}-\frac{1}{\xi}\right)D^2D_\alpha D^\beta\right];\\
\label{gammab}
\Gamma_b[\Phi]&=&\frac{1}{2}\textrm{Tr}\ln\Bigg\{\left[D^2\left(1-\frac{m^2_B+V_3}{\Box}\right)-V_2\right]D_\alpha D^\beta+\left(\frac{m^2_B+V_3}{\Box}-\frac{1}{\alpha}\right)D^2D^\beta D_\alpha\nonumber\\
&-&\left(m+V_1^\prime\right)^2\frac{D^2+m_\Phi+V_0^{\prime\prime}}{\Box-\left(m_\Phi+V_0^{\prime\prime}\right)^2}D_\alpha D^\beta-\left(\frac{V_3}{m_B}\right)^2\frac{D^2D^\beta D_\alpha}{\Box-\alpha m^2_B+\frac{V_3}{m^2_B}\Box}\Bigg\};\\
\label{gammap}
\Gamma_\phi[\Phi]&=&-\frac{1}{2}\textrm{Tr}\ln\left(D^2+m_\Phi+V_0^{\prime\prime}\right).
\eea
Notice that $\omega_\alpha$ and $b_\alpha$ are fermionic variables, so that $\Gamma_\omega$ and $\Gamma_b$ got an overall plus sign.

Now, let us start with the first contribution $\Gamma_\omega$. First, we factor out the inverse of the $\omega_\alpha$-propagator from (\ref{gammao}). Thus, Eq. (\ref{gammao}) can be rewritten as
\bea
\Gamma_\omega&=&\frac{1}{2}\textrm{Tr}\ln\left(D^2D^\gamma D_\alpha-\frac{1}{\xi}D^2D_\alpha D^\gamma\right)+\frac{1}{2}\textrm{Tr}\ln\bigg\{{\delta_\gamma}^\beta+\frac{1}{2\Box}\bigg[\left(-\alpha\frac{m^2_B}{\Box}+\frac{V_3}{m^2_B}\right)D^2D^\beta D_\gamma\nonumber\\
&+&\xi\alpha\frac{m^2_B}{\Box}D^2D_\gamma D^\beta\bigg]\bigg\}.
\eea
Note that the first trace does not depend on the background superfield, then it can be disregarded. The second trace can be split into two parts with the help of the identity $D^\alpha D_\beta D_\alpha=0$. Therefore,
\bea
\label{equation}
\Gamma_\omega=\frac{1}{2}\textrm{Tr}\ln\left\{{\delta_\gamma}^\lambda+\frac{1}{2\Box}\left(-\alpha\frac{m^2_B}{\Box}+\frac{V_3}{m^2_B}\right)D^2D^\lambda D_\gamma\right\}+\frac{1}{2}\textrm{Tr}\ln\left\{{\delta_\lambda}^\beta+\frac{\xi\alpha m^2_B}{2\Box^2}D^2D_\lambda D^\beta\right\}.
\eea
Again, the second trace is a constant independent of the background superfield and it can be dropped. To solve the first trace, we have to perform a series expansion of the logarithm. Therefore,
\bea
\label{expansion1}
\Gamma_\omega&=&-\frac{1}{2}\int d^5z\int\frac{d^3k}{(2\pi)^3}\sum_{n=1}^\infty\frac{1}{n}\left[\frac{1}{2k^2}\left(\alpha\frac{m^2_B}{k^2}+\frac{V_3}{m^2_B}\right)\right]^n(D^2)^nD^{\alpha_2}D_{\alpha_1}D^{\alpha_3}D_{\alpha_2}\cdots D^{\alpha_n}D_{\alpha_{n-1}} \nonumber\\
&\times&D^{\alpha_1}D_{\alpha_n}\delta^2(\theta-\theta^\prime)|_{\theta=\theta^\prime} .
\eea
Each term of the expansion can be evaluated using the $D$-algebra and the following identities:
\bea
\label{delta}
\delta^2(\theta-\theta^\prime)|_{\theta=\theta^\prime}=0 \ ; \ D_\alpha\delta^2(\theta-\theta^\prime)|_{\theta=\theta^\prime}=0  \ ; \ D^2\delta^2(\theta-\theta^\prime)|_{\theta=\theta^\prime}=1.
\eea
Thus, it is possible to show that each term in the expansion (\ref{expansion1}) vanishes. Therefore, we obtain
\bea
\label{finalo}
\Gamma_\omega[\Phi]=0.
\eea
In the context of three-dimensional super-QED, a vanishing contribution to $K^{(1)}(\Phi)$ was also found in Refs. \cite{GNP,Petrov}, where was shown that the contribution of the gauge superfield to $K^{(1)}(\Phi)$ vanishes in the Landau gauge. In contrast to \cite{GNP,Petrov}, we have shown that the contribution of the St\"{u}ckelberg superfield vanishes for any values of the gauge parameters $\alpha$ and $\xi$.

Now, let us consider the contribution of the quantum prepotential $b_\alpha$ to $\Gamma^{(1)}_B[\Phi]$. By repeating the same reasoning that led from (\ref{gammao}) to (\ref{equation}), we can prove that (\ref{gammab}) can be rewritten as
\bea
\label{equation2}
\Gamma_b&=&\frac{1}{2}\textrm{Tr}\ln\left(D^2D_\alpha D^\gamma-\frac{1}{\alpha}D^2D^\gamma D_\alpha\right)+\nonumber\\&+&
\frac{1}{2}\textrm{Tr}\ln\bigg\{{\delta_\gamma}^\lambda+\frac{1}{2\Box}\bigg[D^2\left(\frac{m_B^2+V_3}{\Box}+\frac{(m+V_1^\prime)^2}{\Box-(m_\Phi+V_0^{\prime\prime})^2}\right)+\nonumber\\&+&
V_2+\frac{(m+V_1^\prime)^2(m_\Phi+V_0^{\prime\prime})}{\Box-(m_\Phi+V_0^{\prime\prime})^2}\bigg]D_\gamma D^\lambda\bigg\}+\nonumber\\ &+&
\frac{1}{2}\textrm{Tr}\ln\bigg\{{\delta_\lambda}^\beta-\frac{\alpha}{2\Box^2}\bigg[m_B^2+V_3-\left(\frac{V_3}{m_B}\right)^2
\frac{\Box}{\Box-\alpha m^2_B+\frac{V_3}{m_B^2}\Box}\bigg]D^2D^\beta D_\lambda\bigg\}.
\eea
Notice that only the second trace is nonvanishing and independent of $\alpha$. In order to make progress, we need the identity 
\bea
{\delta_\gamma}^\lambda+AD^2D_\gamma D^\lambda+BD_\gamma D^\lambda=\left({\delta_\gamma}^\alpha+AD^2D_\gamma D^\alpha\right)\left({\delta_\alpha}^\lambda+\frac{B}{1-2\Box A}D_\alpha D^\lambda\right).
\eea
Thus, by applying this identity to (\ref{equation2}), we find
\bea
\Gamma_b&=&\frac{1}{2}\textrm{Tr}\ln\left\{{\delta_\gamma}^\alpha+\frac{1}{2\Box}\left[\frac{m^2_B+V_3}{\Box}+\frac{(m+V_1^\prime)^2}{\Box-(m_\Phi+V_0^{\prime\prime})^2}\right]D^2D_\gamma D^\alpha\right\}\nonumber\\
&+&\frac{1}{2}\textrm{Tr}\ln\bigg\{{\delta_\alpha}^\lambda+\frac{1}{2}\frac{\left[\Box-(m_\Phi+V_0^{\prime\prime})^2\right]V_2+(m+V_1^\prime)^2(m_\Phi+V_0^{\prime\prime})}{\left[\Box-(m_\Phi+V_0^{\prime\prime})^2\right](\Box-m_B^2-V_3)-(m+V_1^\prime)^2\Box}D_\alpha D^\lambda\bigg\}.
\eea
In order to evaluate the second trace (the first one is equal to zero), we shall make the simplifying assumption that $m=V_1=0$. Therefore, under such a simplifying assumption, we find
\bea
\label{expansion2}
\Gamma_b&=&-\frac{1}{2}\int d^5z\int\frac{d^3k}{(2\pi)^3}\sum_{n=1}^\infty\frac{1}{2^n n}\frac{V_2^n}{(k^2+m_B^2+V_3)^n}D_{\alpha_1}D^{\alpha_2}D_{\alpha_2}D^{\alpha_3}\cdots D_{\alpha_{n-1}}D^{\alpha_n}\nonumber\\
&\times&D_{\alpha_n}D^{\alpha_1}\delta^2(\theta-\theta^\prime)|_{\theta=\theta^\prime}.
\eea
Again, with the help of the $D$-algebra and the identities (\ref{delta}), we are able to formally show that
\bea
D_{\alpha_1}D^{\alpha_2}D_{\alpha_2}D^{\alpha_3}\cdots D_{\alpha_{n-1}}D^{\alpha_n}D_{\alpha_n}D^{\alpha_1}\delta^2(\theta-\theta^\prime)|_{\theta=\theta^\prime}=\left\{\begin{array}{rcl}
-2^n(\sqrt{-k^2})^{n-1}, & \mbox{if} & n=2\ell+1\\
0, & \mbox{if} & n=2\ell
\end{array}.
\right.
\eea
Substituting this formula into (\ref{expansion2}), we obtain
\bea
\Gamma_b=\frac{1}{2}\int d^5z\sum_{\ell=0}^\infty\frac{(-1)^{\ell}V_2^{2\ell+1}}{2\ell+1}\int\frac{d^3k}{(2\pi)^3}\frac{(k^2)^\ell}{(k^2+m_B^2+V_3)^{2\ell+1}} \ .
\eea
We can evaluate this well-known integral over the momenta and sum the results over $\ell$ to get
\bea
\label{finalb}
\Gamma_b[\Phi]=-\frac{1}{16\pi}\int d^5zV_2\sqrt{4(m_B^2+V_3)+V^2_2} \ .
\eea
The last (and easiest) contribution which is needed to be calculated is (\ref{gammap}). We can simply repeat the same reasoning that led us to Eqs. (\ref{finalo}) and (\ref{finalb}), but we will not
calculate explicitly $\Gamma_\phi$. Therefore, the final result is given by
\bea
\label{finalp}
\Gamma_\phi[\Phi]=-\frac{1}{16\pi}\int d^5z(m_\Phi+V_0^{\prime\prime})^2 \ .
\eea
Finally, substituting (\ref{finalo}), (\ref{finalb}), and (\ref{finalp}) into (\ref{totalgamma}) and using the relation $\Gamma^{(1)}_B=\int d^5zK^{(1)}_B$, we find
\bea
\label{poteff2}
K^{(1)}_B(\Phi)=-\frac{1}{16\pi}\left[\left(m_\Phi+V^{\prime\prime}_0\right)^2+V_2\sqrt{4(m_B^2+V_3)+V^2_2}\right] \ .
\eea
Similarly to $K^{(1)}_A$ [see Eq. (\ref{poteff1})], $K^{(1)}_B$ is UV finite and, therefore, no additional renormalization is needed. Moreover, $K^{(1)}_B$ is also independent of the gauge-fixing parameters $\alpha$ and $\xi$. In contrast to $K^{(1)}_A$, the functional structure of $K^{(1)}_B$ is not given by a polynomial function of $V^{\prime\prime}_0,V_2$, and $V_3$. In the $\mathcal{N}=1$, $d=3$ superspace, such non-polynomial structure is also found in one-loop effective potentials in the context of higher-derivative theories (see, for example, \cite{GNP}). We conclude this section with the remark that the results (\ref{poteff1}) and (\ref{poteff2}), which were obtained by different methods, coincide with each other when $m=m_B=V_1=V_3=0$. This shows that $K^{(1)}_B$ obtained through evaluation of the matrix trace is consistent with $K^{(1)}_A$ obtained with use of eigenvalues of the mass matrix.

\section{Summary}
\label{Conc}

We formulated a supersymmetric theory of three-dimensional two-form field. In the superfield language, this theory is described by a spinor prepotential $B_{\alpha}$. We started with a gauge invariant strength $G$ defined in terms of $B_{\alpha}$, and further introduced a mass term for this field, a coupling of this field to an usual scalar superfield $\Phi$ and a St\"{u}ckelberg superfield in order to implement gauge symmetry in the presence of the mass term. Afterwards, we calculated the one-loop effective potential of $\Phi$ in a resulting theory, using a functional approach. The effective potential turns out to be finite as it must occur in three-dimensional theories. We explicitly demonstrated that our results are rather analogous to the one-loop results in supergauge theories constructed on the base of the usual vector supermultiplet.

Essentially, the main result of our paper is a first example of successful formulation of a consistent coupling of three-dimensional spinor superfield to  a scalar matter, with the theory turns out to possess gauge symmetry under transformations different from those one in usual supersymmetric QED, and successful calculation of quantum corrections in this theory. Effectively, the main conclusion is that we developed a new supergauge theory with a consistent coupling.

Further development of our study could consist in development of non-Abelian generalization of our theory and in study of higher loop corrections. We expect to do these studies in forthcoming papers.

{\bf Acknowledgments.} Authors are grateful to R. V. Maluf for valuable discussions. P. J. Porf\'{i}rio would like to acknowledge the Brazilian agency CAPES (PDE process number 88881.171759/2018-01) for the financial support. This work was partially supported by Conselho
Nacional de Desenvolvimento Cient\'{\i}fico e Tecnol\'{o}gico (CNPq). The work by A. Yu. P. has been partially supported by the
CNPq project No. 303783/2015-0.

\end{document}